\documentclass[prl,aps,twocolumn,floats,showpacspsfig]{revtex4}
\usepackage{amssymb}

\usepackage{epsfig}
\usepackage{graphicx}
\usepackage{subfigure}

\newcommand{\be}{\begin{equation}}
\newcommand{\ee}{\end{equation}}
\newcommand{\bea}{\begin{eqnarray}}
\newcommand{\eea}{\end{eqnarray}}

\begin{document}

\title{Orbital Dependence of Quasiparticle Lifetimes in $Sr_2RuO_4$}

\author{ R. M. Konik$^1$ and T. M. Rice$^{1,2}$}
\affiliation{$^{1}$Department of  Physics, Brookhaven National Laboratory, Upton, NY 11973-5000, USA\\
$^{2}$Institute f\"ur Theoretische Physik, ETH-Z\"urich, CH-8093  Z\"urich,  Switzerland}
\date{\today}

\begin{abstract}
Using a phenomenological Hamiltonian, we investigate the quasiparticle lifetimes
and dispersions in the three low energy bands, $\gamma$, $\beta$, and $\alpha$ of $Sr_2RuO_4$.  Couplings 
in the Hamiltonian are fixed so as to produce the mass renormalization as measured 
in magneto-oscillation experiments.
We thus find reasonable agreement in all bands between our computed lifetimes and those measured in ARPES
experiments by Kidd et al. \cite{kidd} and Ingle et al. \cite{ingle}. 
In comparing computed to measured quasiparticle dispersions, we however find good agreement
in the $\alpha$-band alone.  
\end{abstract}
\maketitle

\section{Introduction}

The ruthenate $Sr_2RuO_4$ has been extensively studied for a number of reasons, 
chief among them its unconventional superconducting state \cite{mack}. Its electronic 
structure, reviewed in detail by Bergemann et al. \cite{bergemann}, 
has also been of much interest. 
Three bands belonging to the $t_{2g}$  
complex of 4d $Ru$ orbitals cross the Fermi energy. They are divided into two sets.  
One derived from the $d_{xy}$ orbital has a two-dimensional dispersion with little 
dispersion along the c-axis due to the layered structure of the material. 
The second 
set comprises the $d_{xz}$ and $d_{yz}$ bands which have predominantly one-dimensional 
dispersion. A key feature is the absence of hybridization between these sets in a 
single layer due to the opposite parity under reflection about a $RuO_2$ plane. 
This contrast in their dispersion has been invoked by Kidd and collaborators \cite{kidd} 
to explain the strong difference in the energy and temperature dependence of the 
quasiparticle lifetimes between the two sets  observed in recent ARPES (Angle 
Resolved Photoemission Spectroscopy) experiments.  In this note we report on 
some simple calculations to examine this interpretation.

\begin{figure}
\begin{center}
\noindent
\epsfysize=.4\textwidth
\epsfbox{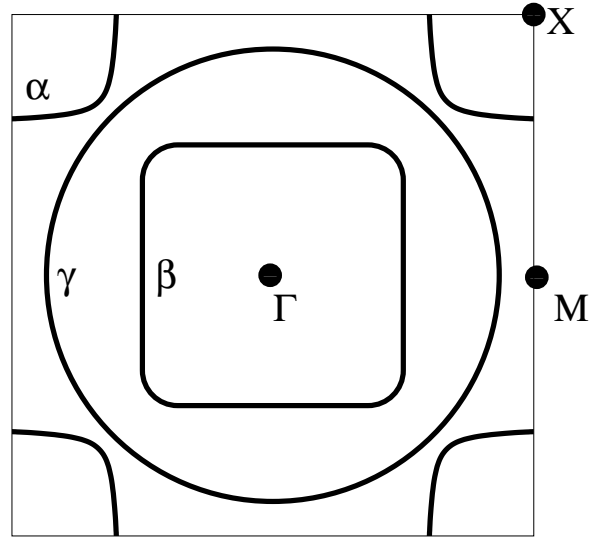}
\end{center}
\caption{A schematic of the three bands on $Sr_2RuO_4$.}
\end{figure}
 
The Fermi surface is illustrated in Fig. 1 and consists of three sheets. The almost 
circular $\gamma$-sheet derives from orbitals with $d_{xy}$ symmetry.  The $\alpha$ 
and $\beta$ sheets 
derive from the approximately linear sections due to the orbitals with $d_{xz}$ and $d_{yz}$ 
symmetry which however hybridize weakly with each other where they cross. 
Kidd and collaborators \cite{kidd} measured the dispersion and linewidths of the 
quasiparticles on the $\gamma$- and the $\beta$-  
sheets along the line $\Gamma$-M  and found a clear difference 
in their behavior as a function of both energy and temperature. 
Ingle et al. \cite{ingle} measured the same quantities for the $\alpha$-band along
the $\Gamma$-X direction finding similar linewidths to those observed
by Kidd et al. in the $\beta$-band.  At first sight this 
difference between the $\gamma$-band and the $\alpha$-, $\beta$-bands
would seem to be simply a consequence of their differing  dispersion and 
the lack of hybridization between them. While direct scattering of a quasiparticle 
between these bands is forbidden due to their different parities, quasiparticles in 
these band will still interact through the Coulomb interactions which leads to 
modifications of their strictly one and two-dimensional character.  To explore this 
effect we undertake low order calculations of the lifetime using phenomenological
interaction strengths.  These interaction strengths are chosen so that the mass renormalization
observed in magneto-oscillation experiments \cite{mack,bergemann} is reproduced.

Doing so we are able to account for the broad features of the observed linewidths.  In particular
we obtain a good fit to the linewidths in the $\gamma-$band as a function of binding energy and we
find, consistent with experiment that the linewidths in the $\beta$- and
$\alpha$-bands are far smaller.  Our computations do however consistently underestimate to a small degree the 
linewidths.  This points to, perhaps, some other contributory mechanism beyond electron-electron interactions.
We also explored the finite temperature behavior of the linewidths.  In both the $\gamma$- and $\beta$- bands,
our computations match well
the observed behaviour seen in \cite{kidd}.  
Unlike the linewidths, the agreement between the measured and computed dispersion relations is mixed.  
For the $\alpha$-band,
good agreement is found while for the $\beta$ and $\gamma$-bands, 
the measured mass renormalization is far less than what 
would be expected from magneto-oscillation experiments.  We comment on this further in the discussion and
conclusion section.

\begin{figure*}
\begin{center}
\noindent
\epsfysize=.4\textwidth
\epsfbox{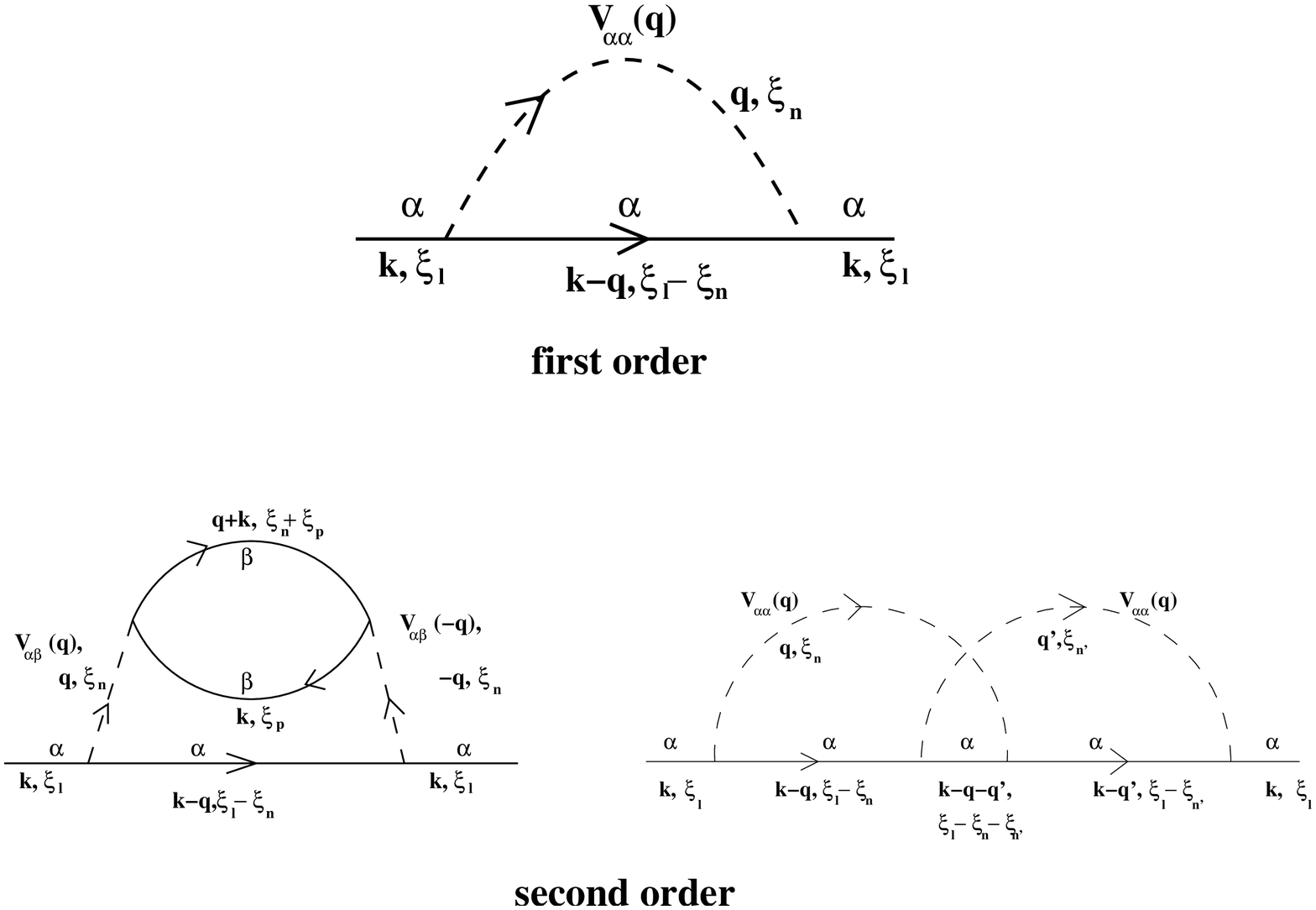}
\end{center}
\caption{Diagrams contributing to the self energy at first and second order.  Here the greek indices $\alpha,\beta$ refer
to generic bands.}
\end{figure*}

\section{Calculations of the Lifetime and Effective Mass}
   
The multi-band Hamiltonian that describes a single layer is as follows
\begin{eqnarray}
{\cal H} &=& {\cal H}_{0} + {\cal H}_{\rm int};\cr
{\cal H}_0 &=& \sum_{k\sigma\mu}\epsilon^{\mu}(k) c^\dagger_{\mu k\sigma}c_{\mu k\sigma};\cr
{\cal H}_{\rm int} &=& \sum_{\mu,\nu,\sigma,\sigma',i}V_{\mu\nu}n_{\mu\sigma}(i)n_{\nu\sigma'}(i).
\end{eqnarray}
Here the greek indices, $\mu,\nu$, are band indices and sum over the three bands, $\alpha,\beta,$ and $\gamma$.
The two-dimensional $\gamma$-band, describing 
orbitals with $d_{xy}$ symmetry, takes the form \cite{morr,ll},
\begin{equation}
\epsilon^\gamma({\bf k}) = -.88\cos(k_x) - .88\cos(k_y) - .56\cos(k_x)\cos(k_y) - .5.
\end{equation}
The one-dimensional $\alpha$- and $\beta$-bands arise from weak hybridization between the $d_{xz}$ and $d_{yz}$ 
symmetry orbitals and appear as
\begin{eqnarray}
\epsilon^{xz}({\bf k}) &=& -.62\cos(a k_x) - .09\cos(a k_y)  \cr 
&& \hskip .5in + .04\cos(a k_x)\cos(a k_y) - .24;\cr
\epsilon^{yz}({\bf k}) &=& -.09\cos(a k_x) - .62\cos(a k_y) \cr
&& \hskip .5in + .04\cos(a k_x)\cos(a k_y) - .24;\cr
\epsilon^{\pm}({\bf k}) &=& \frac{1}{2}(\epsilon^{xz}\pm\epsilon^{yz});\cr
\epsilon^{\alpha/\beta} &=& \epsilon^+ \mp \sqrt{(\epsilon^-)^2+.01}.
\end{eqnarray}
All energies are in eV's and $a=3.86A^o$ is the lattice spacing.
The interaction vertices reduce to four terms describing intraband scattering between
electrons in the $\gamma$-band ($V_{\gamma\gamma}$) and in the 
$\alpha$,$\beta$-bands ($V_{\alpha\alpha}=V_{\beta\beta}$) and interband scattering
between $\gamma$ and the $\alpha,\beta$-bands($V_{\gamma\alpha}=V_{\gamma\beta}$) 
and between the $\alpha$ and $\beta$ bands ($V_{\alpha\beta}$).
Simple estimates of the Fermi-Thomas screening length give 
strong screening due to the large density 
of states at the Fermi energy so we neglect the wavevector dependence of the interaction 
vertices. This leads to four independent parameters which we adjust phenomenologically 
as discussed below.
The calculation of the self energy is restricted to the lowest orders in the 
phenomenological interaction vertices, illustrated in Fig 2. For the lifetime this 
amounts to calculating the decay processes of a single quasihole into three 
quasiparticles with renormalized interaction vertices. These processes will dominate 
the decay rate of low energy quasiholes as multiquasiparticle decays will rise more 
slowly with the energy of the quasihole measured from  the Fermi energy. Our aim 
is to explore the effects of the differing band dispersions on the decay rates. 

Our approach to this problem is then in the spirit of Landau Fermi liquid theory -- we use the necessary 
mass renormalization to
fit the coupling constants (akin to the Fermi liquid parameters) and then determine other quantities 
(the inverse lifetimes)
in terms of these same couplings.  In this spirit, the largeness of certain parameters
(for example. the dimensionless coupling $u_{\gamma\gamma} = K_{F\gamma}V_{\gamma\gamma}/2\pi v_{F\gamma}$),
should not be taken as problematic.  

\renewcommand{\thesubfigure}{}
\renewcommand{\subfigcapskip}{-40pt}
\begin{figure*}
\begin{center}
\vskip -.2in
\subfigure[]{
\epsfysize=1.25\textwidth
\includegraphics[height=7.5cm,angle=-90]{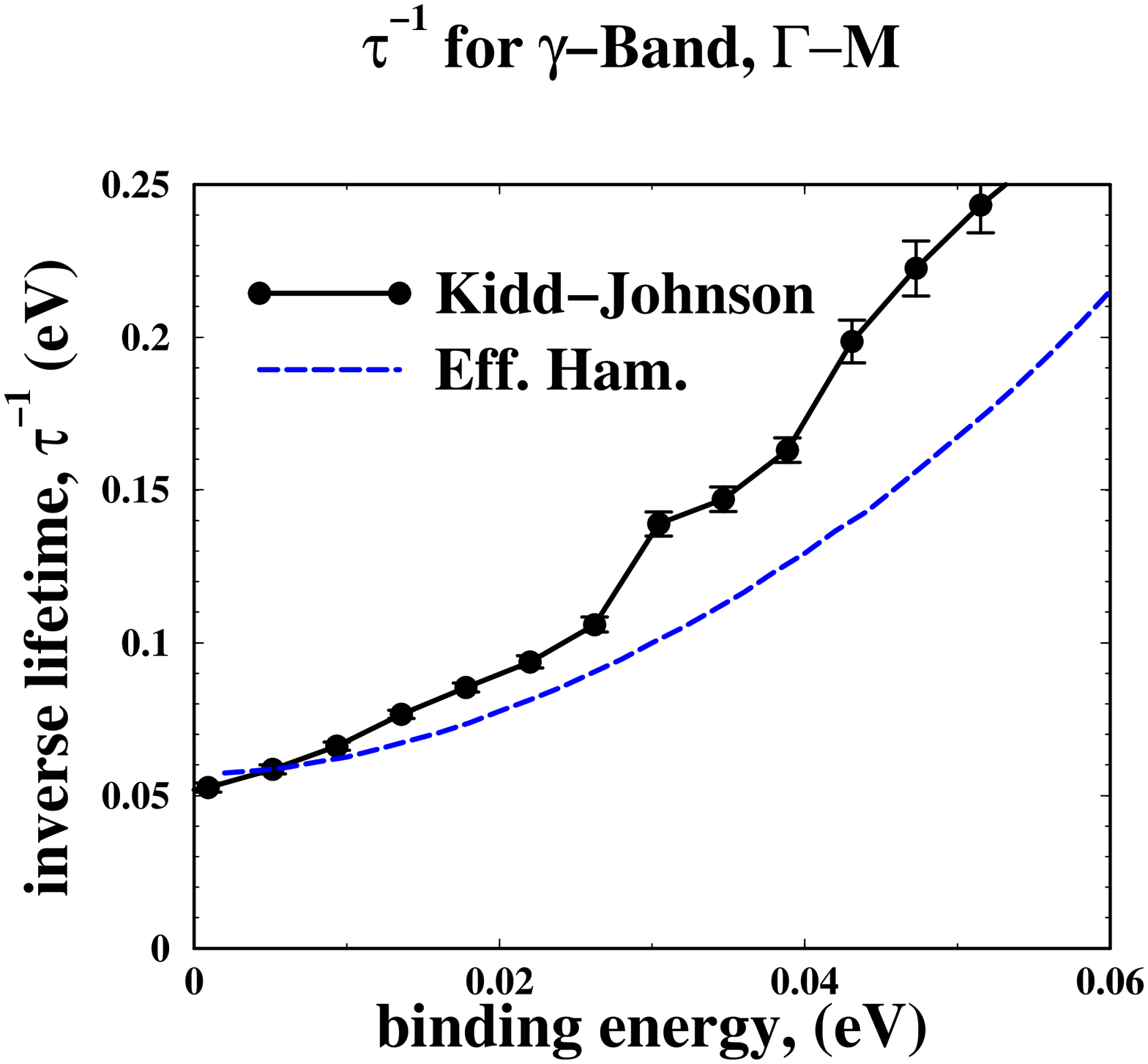}}
\subfigure[]{
\epsfysize=1.25\textwidth
\includegraphics[height=7.5cm,angle=-90]{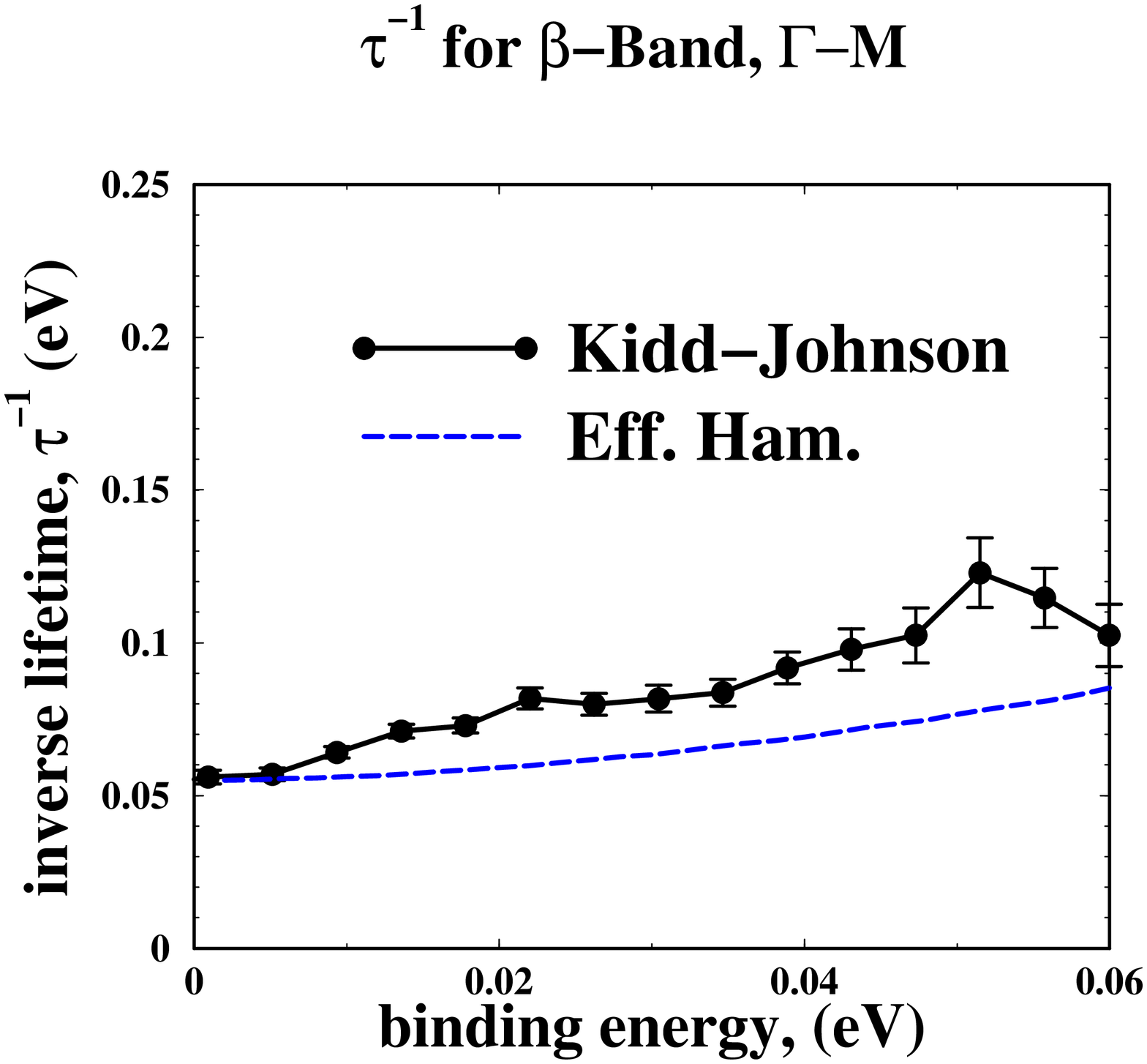}}
\vskip .2in
\end{center}
\centerline{\hbox{}}
\begin{center}
\subfigure[]{
\epsfysize=1.25\textwidth
\includegraphics[height=7.5cm,angle=-90]{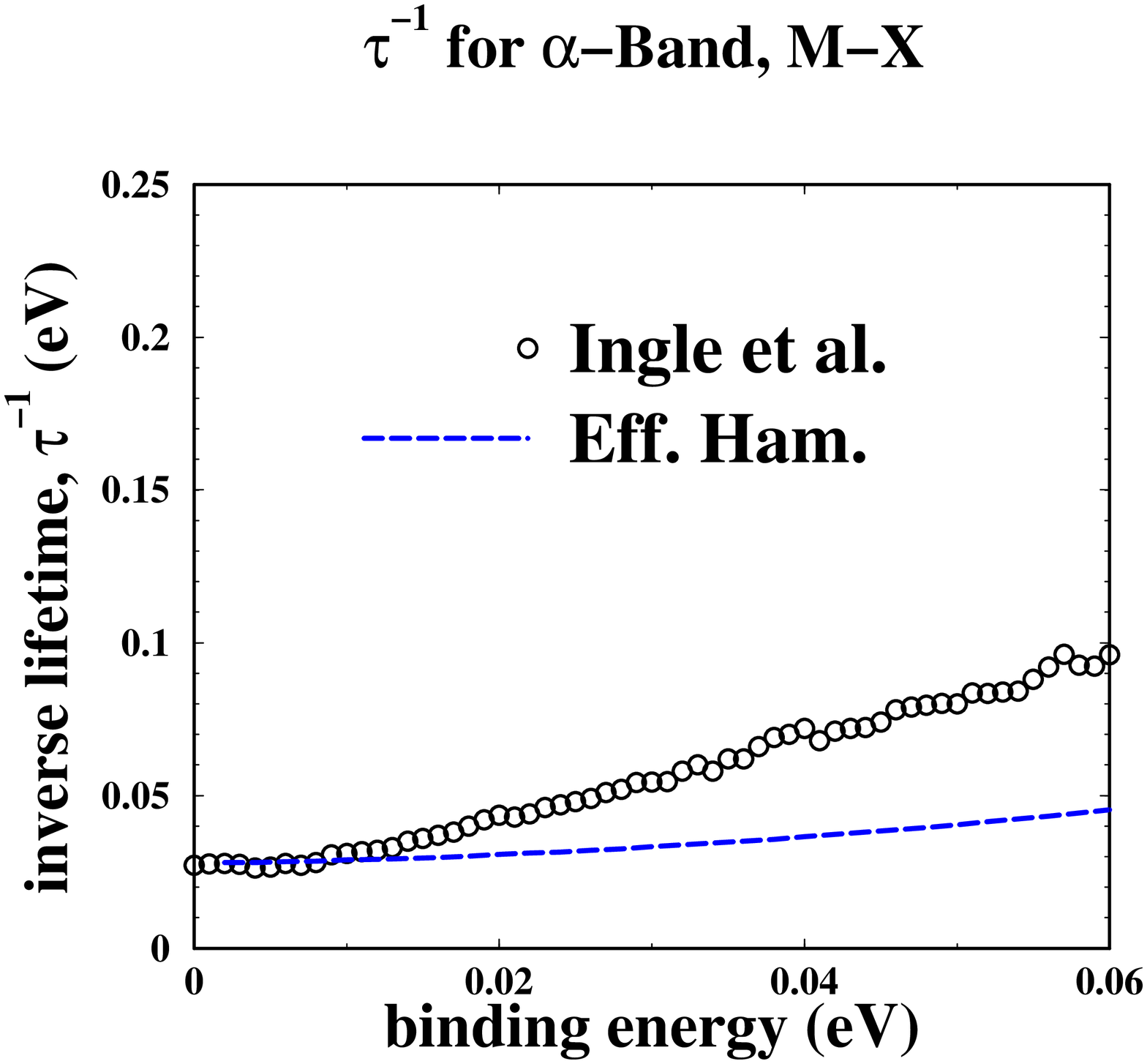}}
\subfigure[]{
\epsfysize=1.25\textwidth
\includegraphics[height=7.5cm,angle=-90]{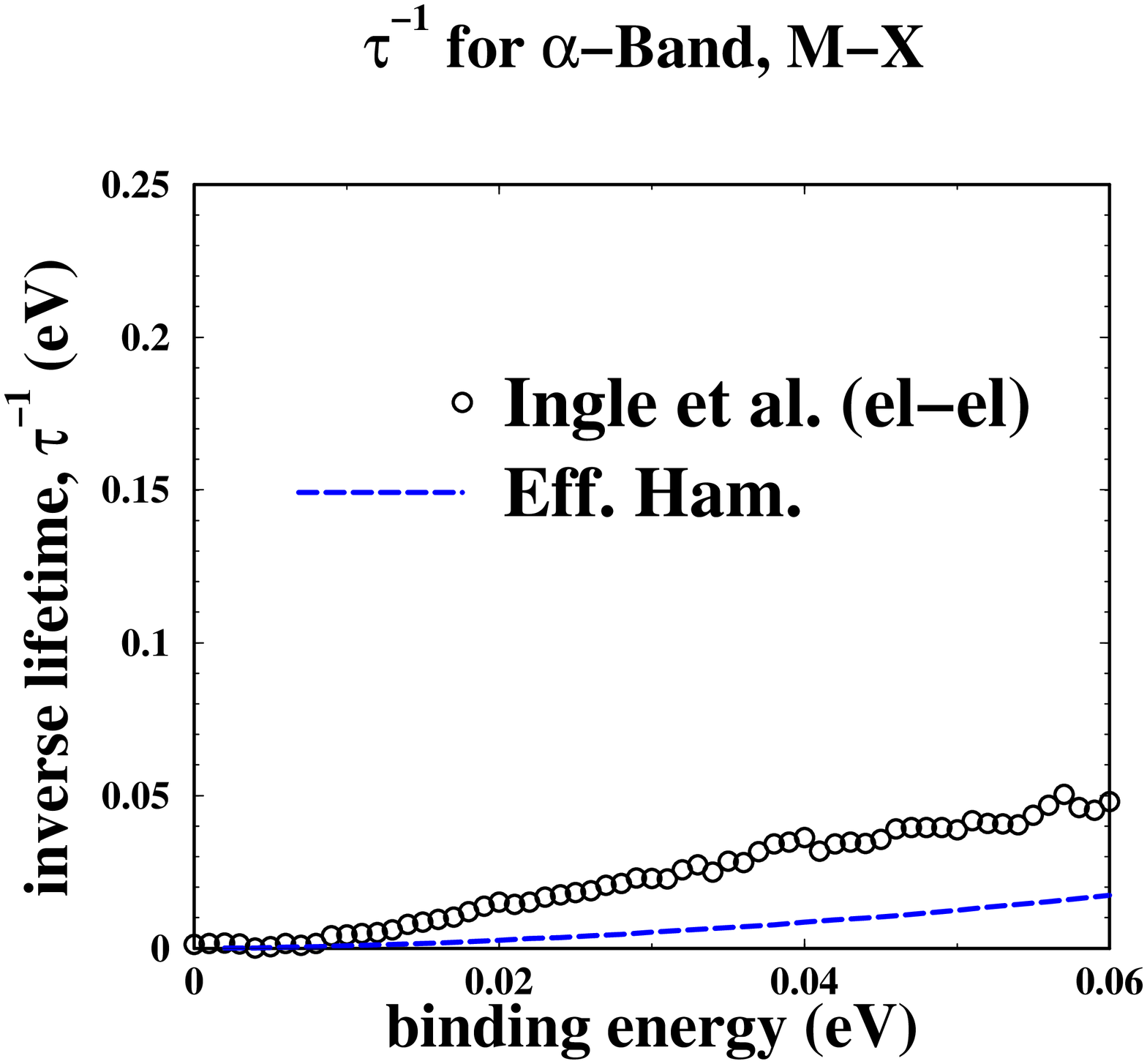}}
\vskip .5in
\end{center}
\caption{Plots of the inverse lifetimes in the three bands, $\gamma$, $\beta$ and $\alpha$.}
\end{figure*}

The real part of the self energy is calculated employing the same set of diagrams.
Any non-zero contribution to the real part at $\omega=0$ at the Fermi surface is 
absorbed into a redefinition of the chemical potential.
The first order diagram, marking the exchange energy, contributes to the real part of 
the self-energy alone.  It is the first in a series of diagrams but is the only member
of the series to make a contribution.  The remaining diagrams in the series simply renormalize
the chemical potential and so are ignored.

\begin{table}
\begin{tabular}{|c|c|c|c|}
\hline
& $\alpha$ & $\beta$ & $\gamma$ \\
\hline
dHvA ($m_*$) & 3.3 & 7.0 & 16 \\
\hline
ARPES & 3.3 & 2.3 & 3.0 \\
\hline
LDA AV ($m_{LDA})$ & 1.2 & 2.3 & 2.3 \\
\hline
$m_R=m_*/m_{LDA}$ & 2.8 & 3.0 & 7.0\\
\hline
\end{tabular}
\caption{Mass renormalization of the electrons in the three bands of $Sr_2RuO_4$ as given by
magneto-oscillation experiments \cite{mack,bergemann} (dHvA), 
ARPES measurements \cite{kidd,ingle}, and LDA computations . 
These three mass renormalizations are all given in terms of the bare electron mass.}
\end{table}

To compute the strengths of phenomenological interaction parameters we insist that
the computed self energy reproduce the mass renormalizations expected from 
magneto-oscillation experiments \cite{mack,bergemann} (see top line of Table 1).
The mass renormalizations induced by the interaction terms in Eq. 2.1
are given by
\begin{equation}
m_R = \frac{m_*}{m_{LDA}} = \frac{1-\partial_\omega{\rm Re}\Sigma(w,k=K_F)}{1+v_{LDA}^{-1}\partial_k{\rm Re}\Sigma(w,k=K_F)}
\end{equation}
where $m_{LDA}$ is the mass renormalization induced by the LDA band structure (also listed in Table I).
The LDA values given in Table I are averages around the bands' Fermi surfaces weighted by the local
density of states (appropriate as the magneto-oscillations masses are themselves weighted averages).  
The interaction strengths are then determined by insisting that $m_R$ equal
$m_*/m_{LDA}$ (as listed in Table 1) at a single point along the Fermi surface of each band.
We do not attempt to average $m_R$ itself around the Fermi surfaces (requiring a computation
of the self-energy everywhere) as too computationally costly.  

In the presence of contact interactions, the $T=0$
self-energy for a given band $\mu$ at first order consists of 
a term independent of both momentum and energy and so may be ignored (at finite $T$, this
term however does do more than renormalize the Fermi energy).  At second order
the contribution takes the form
\begin{eqnarray}
\Sigma_\mu(\omega, k) &=& \sum_\nu \Sigma_{\mu\nu}(\omega, k) \cr\cr
&=& \sum_\nu \frac{V_{\mu\nu}^2}{4\pi^3}\int dq 
\int dx \frac{{\rm Im}\chi_{\nu}(q,x)}{w+i\delta-\epsilon_\mu(k-q)-x}\cr\cr
&& \times (\theta(-x)-\theta(\epsilon_\mu(k-q))),
\end{eqnarray}
where $\chi_\nu$ is the susceptibility of electrons in band $\nu$.
We numerically compute the derivatives of the real part of the self energy necessary
to computing $m_R$ in order
to accurately capture the effect of the dispersion in each of the bands (without
recourse to approximating the bands through a quadratic dispersion relation).

The values of the interaction parameters so determined are given in Table II.  
The first
number marks the value used in later computations of the dispersion and lifetime.
The ranges in brackets mark the range of parameters producing the desired mass
renormalization, $m_R$.  However other choices within the given ranges yield
results for the lifetimes that differ by no more than $10\%$.
\begin{table}
\begin{tabular}{|c|c|c|c|}
\hline
$V_{\gamma\gamma}$ & $V_{\gamma\beta}$ & $V_{\alpha\beta}$ & $V_{\beta\beta}$ \\
\hline
5.9 (5.1-6.5) & 1.4 (0-2) & 1.3 (.8-1.6) & 1.5 (0-2.1) \\
\hline
\end{tabular}
\caption{Strengths of the interaction couplings (all values in $a^2\cdot eV$ where $a$ is
the lattice spacing.}
\end{table}
The largest coupling by far is $V_{\gamma\gamma}$, dictated by the need to produce a mass
renormalization of $m_R \sim 7$.

The real part of the self-energy will have terms beyond linear order in $\omega$ that
however are not necessarily negligible {\it a priori}.  In three dimensions the real
part of the self energy would be expected to take the form 
\begin{equation}
{\rm Re}\Sigma (\omega) = a \omega + b\omega^3 + \cdots.
\end{equation}
By dimensional analysis $b$ would of the form, ${\rm dimensionless~constant}\times E_F^{-2}$,
and so for $\omega$ small, the cubic term can be ignored.
But in two dimensions, the self-energy develops non-analyticities and will appear as
\begin{equation}
{\rm Re}\Sigma (\omega) = a \omega + b\omega^2{\rm sgn}(\omega) + \cdots.
\end{equation}
It thus cannot be immediately neglected.

We can estimate this non-analytic term more precisely.  Using \cite{chubukov},
we know the non-analytic contribution to the self-energy at second order at arbitrary
$\omega$ and $k$ is given for bands with quadratic dispersions \cite{chu1} by
\begin{eqnarray}
{\rm Re}\Sigma^{\rm non-anal.}_{\mu\mu}(\omega,k) &=& \frac{V^2_{\mu\mu}m^2_\mu}{64\pi^2E_{F\mu}}\times\cr
&&\hskip -1in \bigg[(\omega^2+\frac{1}{4}(\omega-\epsilon_\mu(k))^2){\rm sign}(\omega-\epsilon_\mu(k))\cr
&&\hskip -1in + (\omega^2-\frac{1}{4}(\omega-\epsilon_\mu(k))^2){\rm sign}(\omega+\epsilon_\mu(k))\cr
&&\hskip -1in + \omega^2{\rm sign}(\omega) + \omega^2{\rm sign}(\frac{\omega}{\epsilon_\mu(k)});\cr
&&\hskip -1in - (\omega+\epsilon_\mu(k))^2{\rm sign}(\frac{\omega+\epsilon_\mu(k)}{\epsilon_\mu(k)})\bigg];\cr\cr
{\rm Re}\Sigma^{\rm non-anal.}_{\mu\neq\nu}(\omega,k) &=& \frac{V^2_{\mu\nu}m_\mu m_\nu^2}{32\pi^2K_{F\mu}K_{F\nu}}\times\cr
&&\hskip -1in \bigg[(\omega^2+\frac{1}{4}(\omega-\epsilon_\mu(k))^2){\rm sign}(\omega-\epsilon_\mu(k))\cr
&&\hskip -1in + (\omega^2-\frac{1}{4}(\omega-\epsilon_\mu(k))^2){\rm sign}(\omega+\epsilon_\mu(k))\bigg],
\end{eqnarray}
where $E_{F\mu}$ is the effective bandwidth of band $\mu$, and $m_\mu = K^2_{F\mu}/2E_{F\mu}$ with
$K_{F\mu}$ the band's corresponding Fermi wavevector.
$\Sigma^{\rm non-anal.}_{\mu\mu}(\omega,k)$ has two contributions: one arising
from the non-analyticity in the electron polarizability, $\chi_\mu (\omega,q)$, at
$q$ near 0 and one from another non-analyticity in $q$ near $2K_{F\mu}$ \cite{chubukov}.
In contrast, the interband contribution to the self energy, $\Sigma_{\mu\nu}$,
only sees a non-analytic contribution from $\chi_\nu (\omega, q\sim 0)$.  To estimate the contribution these
terms make to the self-energy we need to estimate the appropriate values for the effective (quadratic)
masses appearing in the above.  To do so we write down the corresponding expressions for the imaginary
part of the self energy:
\renewcommand{\thesubfigure}{}
\renewcommand{\subfigcapskip}{-40pt}
\begin{figure*}
\begin{center}
\subfigure[]{
\epsfysize=1.35\textwidth
\includegraphics[height=5.75cm,angle=-90]{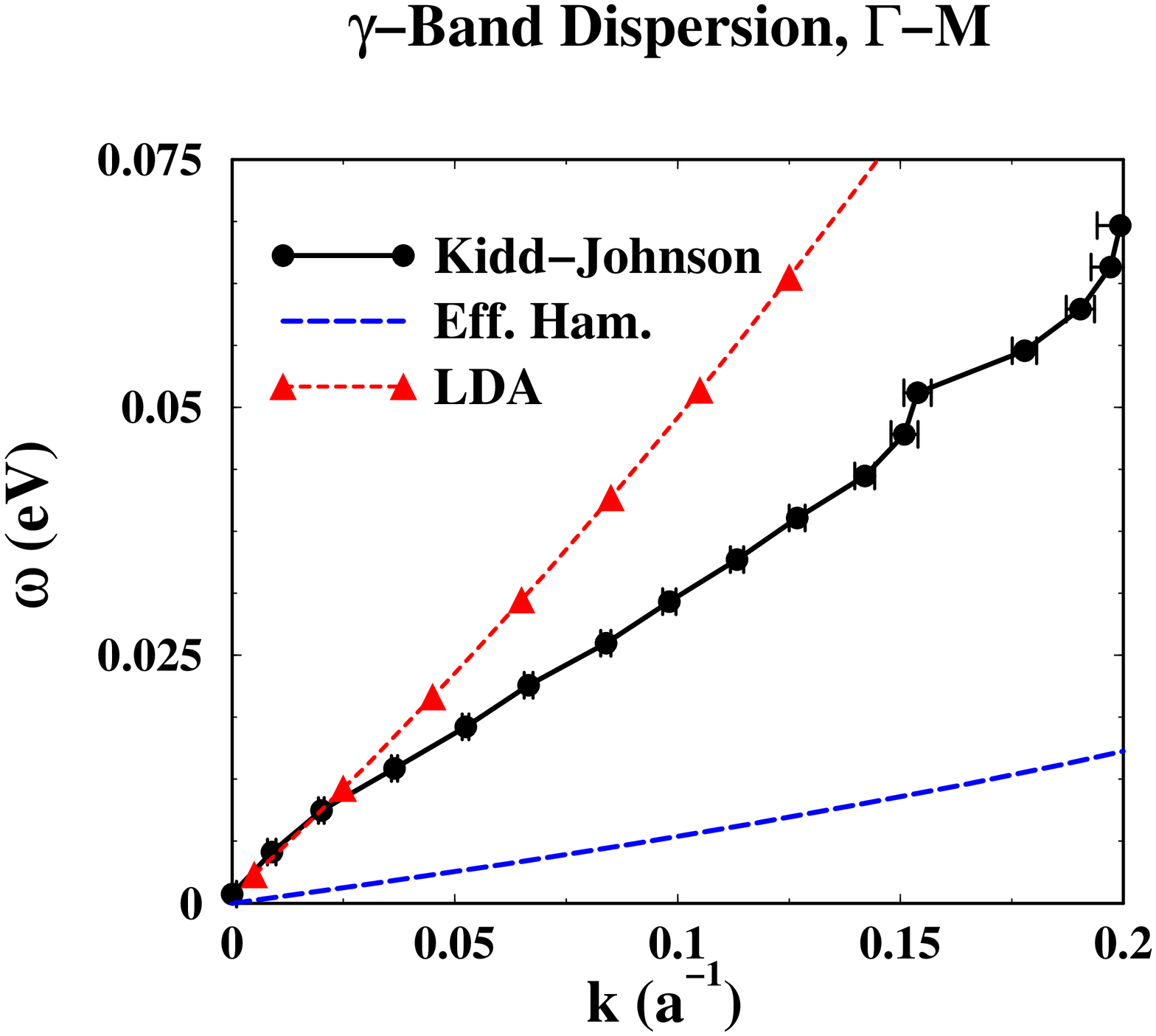}}
\subfigure[]{
\epsfysize=1.35\textwidth
\includegraphics[height=5.75cm,angle=-90]{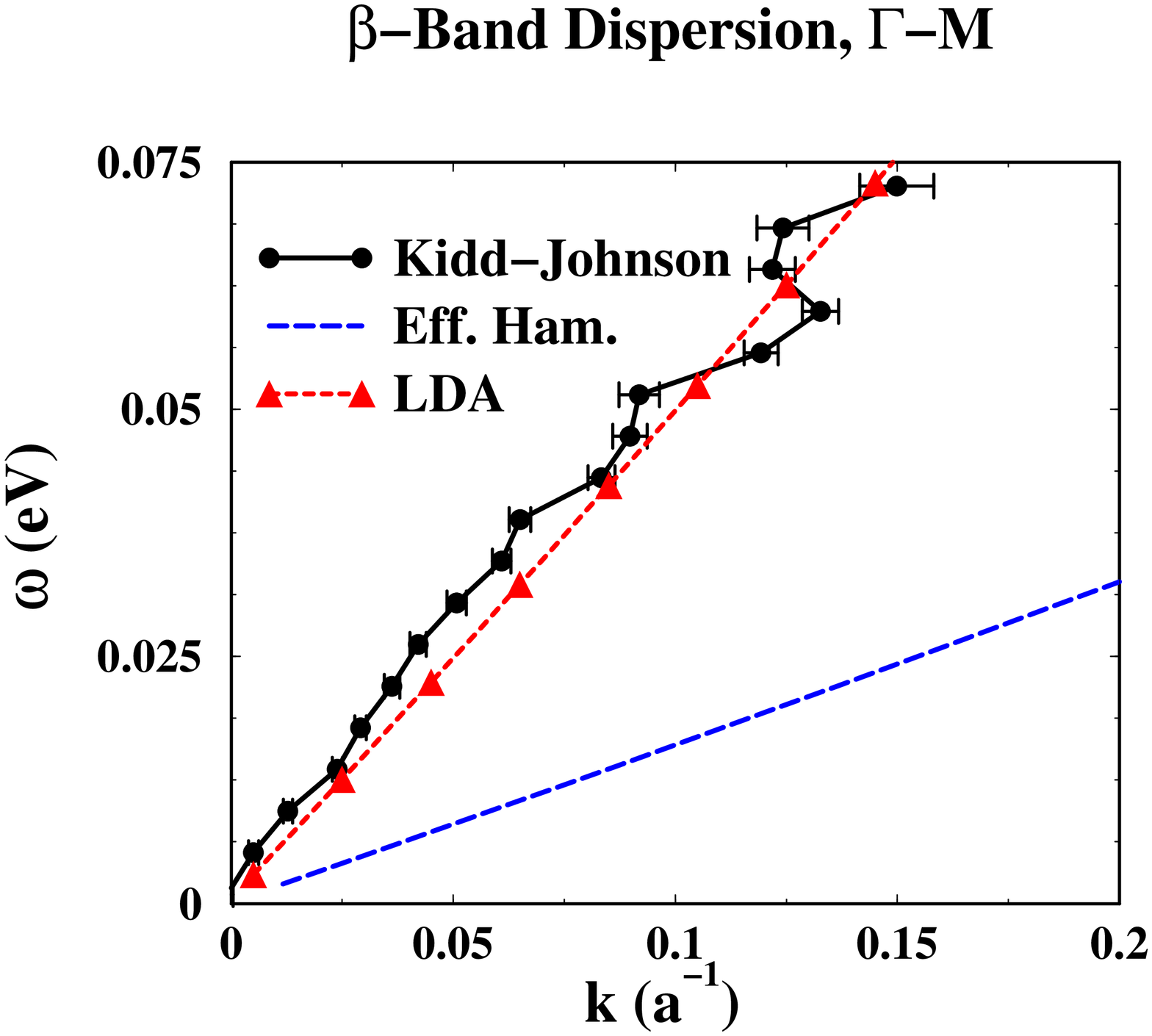}}
\subfigure[]{
\epsfysize=1.35\textwidth
\includegraphics[height=5.75cm,angle=-90]{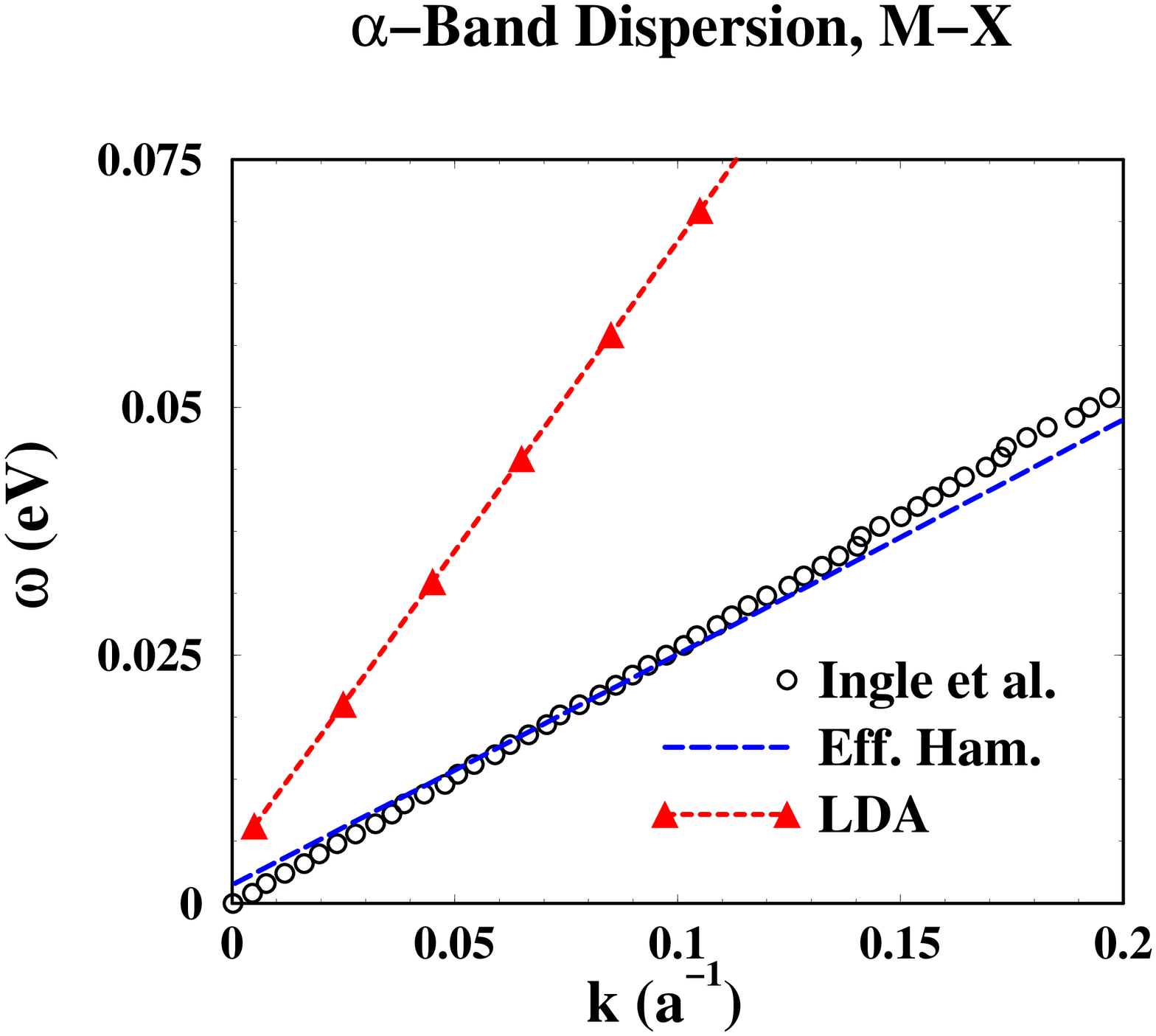}}
\end{center}
\vskip .2in
\caption{Plots of the dispersion in the three bands, $\gamma$, $\beta$ and $\alpha$.}
\end{figure*}
\begin{eqnarray}
{\rm Im}\Sigma_{\mu\mu}(\omega,k) &=& \frac{V^2_{\mu\mu}m^2_\mu}{32\pi^3 E_\mu}\times\cr
&&\hskip -1.2in \bigg[(\omega^2+\frac{1}{4}(\omega-\epsilon_\mu(k))^2)\ln(\big|\frac{\omega-\epsilon_\mu(k)}{E_{F\mu}}\big|)\cr
&&\hskip -1.2in + (\omega^2-\frac{1}{4}(\omega-\epsilon_\mu(k))^2)\ln(\big|\frac{\omega+\epsilon_\mu(k)}{E_{F\mu}}\big|)\cr
&&\hskip -1.2in - \omega^2 + \frac{\omega\epsilon_\mu(k)}{2}\cr
&&\hskip -1.2in + \omega^2\ln(\big|\frac{\omega}{E_{F\mu}}\big|) + \omega^2\ln(\big|\frac{\omega}{\epsilon_\mu(k)}\big|);\cr
&&\hskip -1.2in - (\omega+\epsilon_\mu(k))^2\ln(\big|\frac{\omega+\epsilon_\mu(k)}{\epsilon_\mu(k)}\big|)\bigg];\cr\cr
{\rm Im}\Sigma_{\mu\neq\nu}(\omega,k) &=& \frac{V^2_{\mu\nu}m_\mu m_\nu^2}{16\pi^3 K_{F\mu}K_{F\nu}}\times\cr
&&\hskip -1.2in \bigg[(\omega^2+\frac{1}{4}(\omega-\epsilon_\mu(k))^2)\ln(\big|\frac{\omega-\epsilon_\mu(k)}{E_{F\mu}}\big|)\cr
&&\hskip -1.2in + (\omega^2-\frac{1}{4}(\omega-\epsilon_\mu(k))^2)\ln(\big|\frac{\omega+\epsilon_\mu(k)}{E_{F\mu}}\big|) \cr
&&\hskip -1.2in - \omega^2 + \frac{\omega\epsilon_\mu(k)}{2}\bigg].
\end{eqnarray}
To determine $E_{F\mu}$, we fit a numerical evaluation of 
the imaginary part of self-energy via Eqn. 2.5 to the above expression.  (The non-analytic portion
of ${\rm Im}\Sigma(\omega,k)$ dominates because of the presence of the logs.)  Having so extracted
effective Fermi energies, $E_{F\mu}$ (see Table III), we then are able to evaluate the non-analytic real part of the self-energy.
We find that it makes a significant contribution only for $\omega > .04eV$ and then only for
${\rm Re}\Sigma_{\gamma\gamma}$ -- the dispersion relations for the $\alpha$ and $\beta$-bands are
insensitive to this correction.

\begin{table}
\begin{tabular}{|c|c|c|}
\hline
$E_{F\gamma}$ & $E_{F\beta}$ & $E_{F\alpha}$ \\
\hline
0.6 & 0.5 & 1. \\
\hline
\end{tabular}
\caption{The effective bandwidths for each of the three bands.}
\end{table}

While the spirit of our computation is Landau Fermi liquid theory, 
we can nonetheless address the question of what contribution higher order diagrams (beyond those in Figure 2)
make to the inverse lifetime.  This is a difficult question of course, but for at least an infinite subset of diagrams studied by Chubukov et al. \cite{chubukov}, we can give a definitive answer: not much.
We focus on $\Sigma_{\gamma\gamma}$ (i.e. the contribution to the self energy
of the $\gamma$-band due to the polarizability of $\gamma$-band electrons) as $V_{\gamma\gamma}$ is the largest coupling.
Chubukov et al. \cite{chubukov} have studied the behaviour of higher order forward scattering terms in the 
$V_{\gamma\gamma}$-perturbative series.  
They found that the relevant expansion parameter
is $s = u_{\gamma\gamma}^2\omega/(\omega-\epsilon_\gamma(k))$.  
In terms of our lifetime computations, we are far off the mass shell because of the
large mass renormalizations involved, i.e. $(\omega-\epsilon_\gamma(k))/\omega \sim 6$, $s$ will be smaller than might first appear,
i.e. $s \sim 4$.  While the radius of convergence of the series is unclear, if the series needs to be resummed, the effects are
not particularly drastic.  The terms in ${\rm Im}\Sigma (\omega , k)$ (see Eqn. 2.8) behaving as $\omega^2\ln (\omega)$ are modified to 
\begin{equation}
\omega^2\ln (\frac{\omega}{E_{F\gamma}}) \rightarrow \omega^2 (\frac{1}{2}\ln (\frac{\omega}{E_{F\gamma}}) - \frac{1}{2}|\ln (u^2)|)
\end{equation}
Given that $u_{\gamma\gamma} \sim 5$ and $E_{F\gamma} \sim 0.5eV$, 
for very small $\omega$ we would then expect higher order contributions to lead to a reduction in the
inverse lifetime while for omega larger than $.02eV$, higher order contributions leave the answer effectively unchanged.

\renewcommand{\thesubfigure}{}
\renewcommand{\subfigcapskip}{-40pt}
\begin{figure*}
\begin{center}
\vskip -.2in
\subfigure[]{
\epsfysize=1.25\textwidth
\includegraphics[height=6cm,angle=-90]{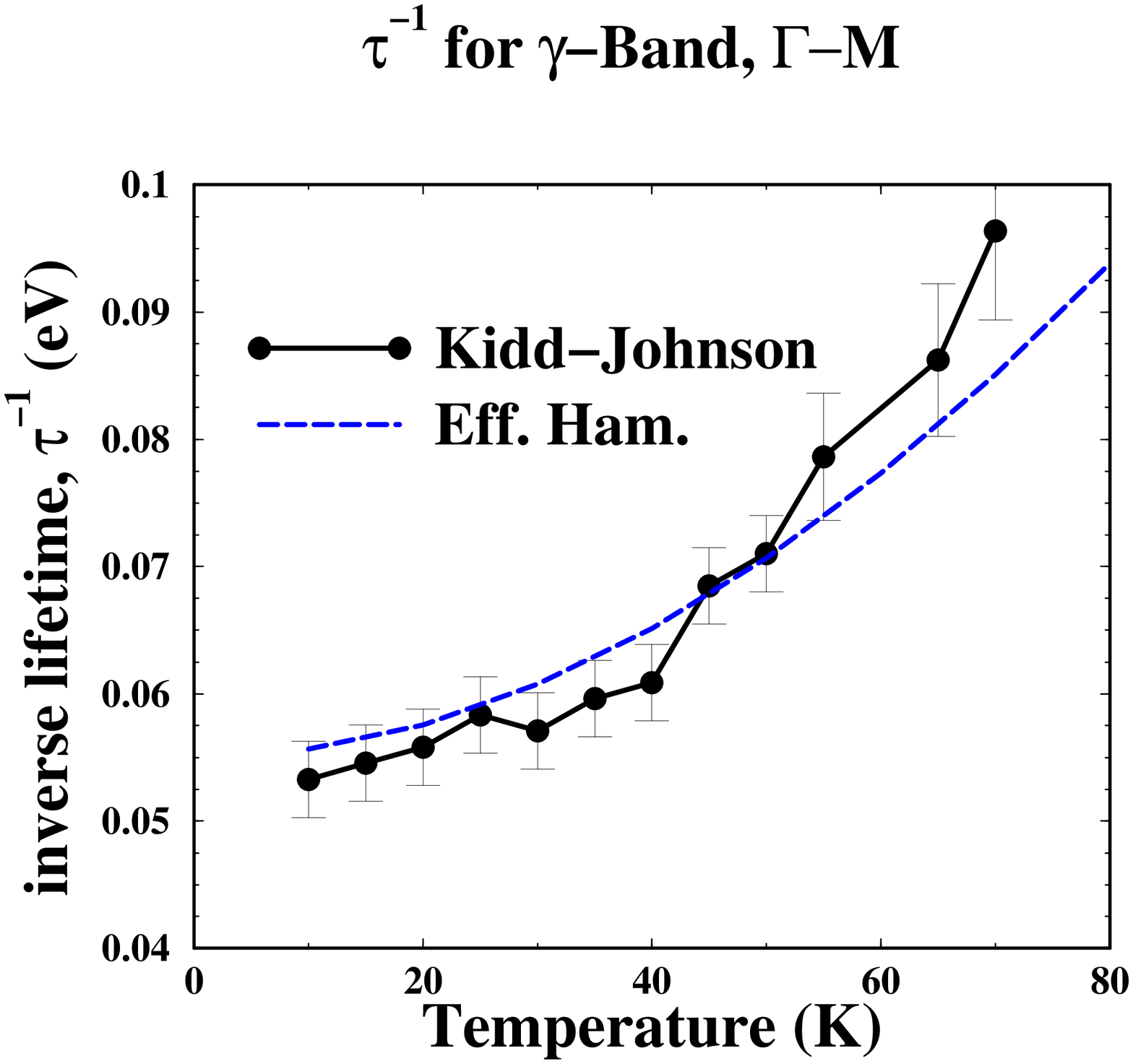}}
\subfigure[]{
\epsfysize=1.25\textwidth
\includegraphics[height=6cm,angle=-90]{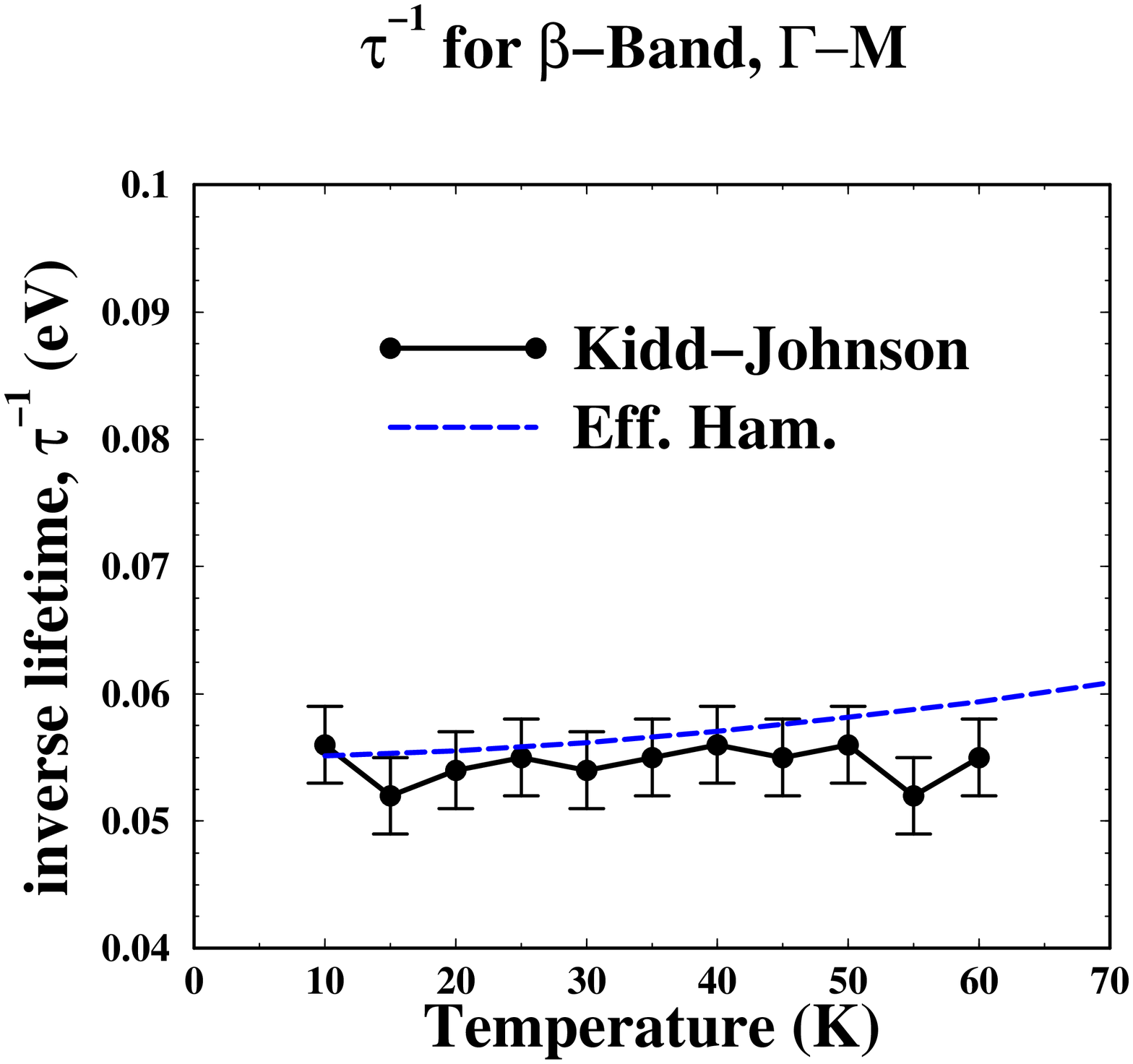}}
\end{center}
\vskip .2in
\caption{Plots of the temperature dependence in the $\gamma$- and $\beta$-bands..}
\end{figure*}
  
\section{Comparison to ARPES Results}

The ARPES studies of Kidd and collaborators \cite{kidd} reported inverse lifetime results 
for a hole in the $\gamma$ and $\beta$ 
bands measured along the $\Gamma-M$ direction as a function of energy 
and of temperature.  There is a pronounced difference 
between the bands with the inverse lifetime for holes in the $\gamma$-band 
rising much more
rapidly with energy and temperature than in the $\beta$-band.  The results for the
$\beta$-band are mimicked by the linewidths observed in the $\alpha$-band in \cite{ingle}.

In the top panel of Figure 3, we plot our computed inverse lifetime as a function of binding
energy for the $\gamma$ and $\beta$-bands along $\Gamma-M$.  To our computed result of the $\gamma$-band,
we have added a static impurity contribution of ${\rm Im}\Sigma_{\rm imp}=.055eV$.  This is chosen so that
the experimental and theoretical values of the inverse lifetime coincide at $\omega=0$.  To arrive
at the experimental values of the inverse lifetime, we have taken the MDC widths as measured by
Kidd et al. and multiplied them by the corresponding LDA value of the band velocity, $v_{F-LDA}$.  For the $\gamma$-band
this velocity changes rapidly near the Fermi surface and so we have correspondingly employed a $k$-dependent
$v_{F-LDA}$.  This yields a slightly
larger experimental inverse lifetime than reported in \cite{kidd}.
While our computed values undershoot the measured values of the inverse lifetime, the agreement 
is reasonable for the $\gamma$-band and would be improved significantly if we treated ${\rm Im}\Sigma_{\rm imp}$
as a completely free parameter.  In particular, we note that
the uncertainties in the measured values do not reflect the uncertainty in the correct value of the
bare LDA velocity to employ in converting an MDC width to an inverse lifetime.

For the $\beta$-band the computed and measured values of the inverse lifetime 
broadly match. In particular, the predicted and measured linewidths are much smaller
than in the $\gamma$-band.   Again we have added an impurity contribution
to the computed inverse lifetime values so that the measured and computed values agree
at zero binding energy.  We do note however that the computed inverse lifetime is smaller
than that measured.  Over a range of binding energies of $.06eV$, the
measured inverse lifetime increases by $\sim .05eV$ whereas the computed value
increases only by half that.  This disagreement may be in part explained by the presence at larger energies 
of correspondingly large error bars.  However, unlike for the $\gamma-$ band, the band velocity 
in the $\beta$-band is not a strong function of wavevector.

In the bottom two panels of Figure 3, we plot our computed lifetime for the $\alpha$-band along
the $M-X$ direction vs that measured by Ingle et al. \cite{ingle}.
Ingle et al. use samples with freshly cleaved surfaces where the $\alpha$-band
can be resolved without employing surface aging.
In the left panel, we plot our computations, again with a correcting 
${\rm Im}\Sigma_{\rm imp}$, against the values measured by Ingle et al..  In the right
panel we plot our computations (with no impurity correction) against the measured electron-electron
contribution to the self-energy.  Ingle et al. deduce this contribution by subtracting out 
an estimate of both the phonon contribution and the impurity contribution.
Again we see that the computed value of the inverse lifetime in the $\alpha$-band is much
less than that of the $\gamma$-band.  However like the $\beta$-band, this value is less than that
measured.  
If we look at only the value imputed to the electron-electron contribution, we see a rise
in the inverse lifetime of $.05eV$ over a $.06eV$ range of binding energies whereas
our computed value rises by a little less than $.02eV$.  

In Figure 4 we compare the corresponding computed and measured dispersion relations for the three bands.
The computed dispersions are essentially straight lines whose slope is determined by 
the value of the dHvA mass renormalizations (as this is how we determine the coupling strength).  
We have also plotted the LDA dispersion relations.  
For the $\gamma$- and $\beta$- bands, there is strong disagreement between measured and computed
values.  Alternatively, the measured dispersion for the $\beta$- and $\gamma$- bands show
a mass renormalization far smaller than seen in magneto-oscillation experiments.  For the $\beta$-band,
the measured mass renormalization equals its LDA prediction.  In the $\alpha$-band,
in contrast, the dispersion closely matches the computed prediction, that is, the 
mass renormalization in the $\alpha$-band measured by ARPES closely matches that
measured by dHvA.

Finally in Figure 5 we compare the computed temperature dependencies of the inverse
lifetimes at zero binding energy in the $\gamma$- and $\beta$-bands to that measured by
Kidd et al. \cite{kidd}.  The $\beta$-band results are well-matched.  Both computed and measured
values show only a weak dependence on temperature over a 100K range.  
Our computations, nonetheless are consistent with a quadratic Fermi-liquid like dependency on the temperature.

The $\gamma$-band results also agree well with that reported in \cite{kidd}.  We roughly
expect the temperature dependency to behave as $A_T T^2\log (E_F/T)$ where $A_T$ is some constant.  Similarly
we expect the dependency of $\tau^{-1}$ on binding energy to be (at leading order,
modulo the complications of Eqn. 9) $A_\omega \omega^2 \log (E_F/\omega)$.
From fits of the two computed curves we find that $A_T/A_w \sim 10$.  The value of this ratio compares well
with the expected value, $\pi^2$, from a single quadratic band \cite{chubukov}.

\section{Discussion and Conclusions}

A primary conclusion from our analysis to the inverse 
lifetimes and effective masses is that there is a marked difference between the two sets of 
bands, $\gamma$, and $\alpha$ and $\beta$ in terms of effective coupling
strengths.  This difference is not a 
simple consequence of the band dispersion and its effect 
on the density 
of states in intraband scattering processes, but rather it is due to a stronger 
effective intraband interaction in the $\gamma$-band at low energies. This conclusion 
is rather surprising since all three bands derive from orbitals belonging to the 
same $t_{2g}$ manifold and standard multiband Hubbard models assign equal onsite interactions 
to all three orbitals \cite{eremin,millis}. 
Our results point to substantially different renormalizations 
of the low energy effective intraband interactions in the two sets of bands.

The different standing of the band $\gamma$ from the bands $\alpha$ and $\beta$ may
arise from the latter two band's highly one dimensional character.  This character
raises the question of whether some form of Luttinger liquid behavior is occurring. The 
fact that the transport properties show clear three-dimensional Landau-Fermi 
behavior \cite{bergemann} 
shows this behavior dominates at low energy.  Nonetheless a crossover to Luttinger liquid 
behavior could appear at higher energies. However the ARPES data reported by both Kidd 
et al. \cite{kidd} and Ingle et al. \cite{ingle} show no sign of such a crossover up to energies of 60 meV. 
Both the $\beta$ and $\alpha$ bands appear to have inverse lifetimes that depend
quadratically on the binding energy.
This perhaps is due to the hybridization and interband scattering between the three bands
which act to suppress a strictly one-dimensional character in the $\alpha,\beta$-
bands.

Our computed inverse lifetimes  broadly match the scale of the observed linewidths in all three bands.
Nonetheless they also consistently underestimate the corresponding measured values, more
pronouncedly in the $\alpha$- and $\beta$- bands than in the
the $\gamma$-band.  This might suggest that some additional mechanism is making a contribution
to the self energy, at least in the $\alpha$- and $\beta$-bands.  This might point to a possible role
played by some bosonic mechanism such as phonons.  However if the shortfall is to be explained by phonons,
phonons need to make a greater contribution to the self energy than posited in Ingle et al. \cite{ingle}.
As we can see from the bottom right panel of Figure 2, the portion of the self-energy in the $\alpha$-band
measured by \cite{ingle} due to electron-electron interactions is double that computed in our
effective model.

In the $\gamma$- and $\beta$-bands we see that ARPES predicts a mass renormalization far
smaller than that found in magneto-oscillation experiments.  This might suggest that there is an
effective scale in the problem below the sensitivity of typically ARPES measurements (i.e. $\ll 1 meV$).
However we also found that the ARPES dispersion in the $\alpha$-band closely matches that predicted
by dHvA measurements.  This might suggest that the surface aging performed by \cite{kidd}
to distinguish the bulk $\gamma$- and $\beta-$ bands from the surface counterparts changes the mass 
renormalization in some 
unexpected fashion.  However our match to the $\alpha$-band dispersion measured in \cite{ingle}
is not without difficultly.  Our match to their dispersion is predicated solely on electron-electron
interactions.  If we were to ascribe a role to phonons, it would mean we have overestimated the coupling
strengths, $V_{\mu\nu}$.   This in turn would mean we have overestimated the inverse lifetimes.
But our inverse lifetimes are already smaller than the measured values.  It is unclear whether
phonons could then self-consistently make up the difference.  Nor is it unproblematic to have phonons
be the dominant interaction in a material where it is believed electron-electron interactions
are responsible for its superconductivity.

Of course, the discrepancies found in regards to
the real part of self energy might simply point to a need to go beyond our use of low order 
diagrams based on 
an effective Hamiltonian.  It would be interesting to attempt a more sophisticated treatment of the set 
of interacting
bands in $Sr_2RuO_4$.  One approach may be to adopt a functional RG approach \cite{shankar}.  Extensive
work of this sort has already been done on one-band two dimensional systems \cite{honer}.  
It should be possible to extend this work to a multi-band system.

We thank both Peter Johnson and Tim Kidd for numerous useful
discussions and direct access to their data.  We are also grateful
to A. Damascelli for providing us with data from Ref. \cite{ingle}.
RMK acknowledges support
from US DOE under contract number DE-AC02 -98 CH 10886. TMR 
acknowledges hospitality from the Institute for Strongly
Correlated and Complex Systems at BNL. 



\begin{thebibliography}{99}

\bibitem{kidd}
T. E. Kidd, T. Valla, A. V. Fedorov, P. D. Johnson, R. J. Cava and M. K. Haas,
Phys. Rev. Lett. {\bf 94}, 107003, (2005).

\bibitem{ingle}
N. J. C. Ingle, K. M. Shen, F. Baumberger, W. Meervasana, D. H. Lu, Z. X. Shen, 
A. Damascelli, S. Nakatsuji, Z. Q. Mao, Y. Maeno, T. Kimura and Y. Tokura,
Phys. Rev. B {\bf 72}, 205114, (2005). 

\bibitem{mack}
A. P. Mackenzie and Y. Maeno, Rev. Mod. Phys. {\bf 75} 657 (2003).

\bibitem{bergemann}
C. Bergemann, A. P. Mackenzie, S. Julian, D. Fortsythe and E. Ohmichi, 
Adv. Phys. {\bf 52} 639 (2003).

\bibitem{morr}
D. Morr, P. Trautman, and M. Graf, Phys. Rev. Lett. 86 (2001) 5978.

\bibitem{ll}
A. Liebsch and A. Lichtenstein, Phys. Rev. Lett. 84 (2000) 1591.

\bibitem{eremin}
I. Eremin, D. Manske and K. H. Bennemann, Phys. Rev. B {\bf 65}, 220502(R) (2002).

\bibitem{millis}
S. Okamoto and A. Millis, cond-mat/0402267.

\bibitem{chubukov} 
A. Chubukov and D. Maslov, Phys. Rev. B {\bf 68}, 155113 (2003);
A. Chubukov, D. Maslov, S. Gangadharaiah, L. Glazman, Phys. Rev. B {\bf 71}, 205112 (2005).

\bibitem{chu1}
Deviations from a quadratically dispersing band can lead to a change in
the non-analytical structure of the self-energy (A. Chubukov and A. Millis, cond-mat/0604496)
if the curvature of the band at the Fermi surface in the transverse direction vanishes.
This is not the case here.

\bibitem{shankar}
R. Shankar, Rev. Mod. Phys. {\bf 66}, 129 (1994).

\bibitem{honer}
C. Honerkamp, M. Salmhofer, and T. M. Rice, Euro. Phys. J. B {\bf 27}, 127 (2004),
C. Honerkamp, D. Rohe, S. Andergassen, and T. Enss, Phys. Rev. B {\bf 70}, 235115 (2004).

\end{thebibliography}
\end{document}